\documentclass{Interspeech}

\interspeechcameraready 

\usepackage[caption=false, position=top]{subfig}

\title{Articulatory Feature Prediction from Surface EMG during Speech Production}

\author[affiliation={1}]{Jihwan}{Lee}
\author[affiliation={1}]{Kevin}{Huang}
\author[affiliation={1}]{Kleanthis}{Avramidis}
\author[affiliation={2}]{Simon}{Pistrosch}
\author[affiliation={2}]{Monica}{Gonzalez-Machorro}
\author[affiliation={1}]{Yoonjeong}{Lee}
\author[affiliation={2}]{Björn}{Schuller}
\author[affiliation={3}]{Louis}{Goldstein}
\author[affiliation={1}]{Shrikanth}{Narayanan}

\affiliation{Signal Analysis and Interpretation Laboratory}{University of Southern California}{USA}
\affiliation{CHI -- Chair of Health Informatics} {Technical University of Munich}{Germany}
\affiliation{Department of Linguistics}{University of Southern California}{USA}

\email{\{jihwan, kevinyhu, avramidi, yoonjeol, louisgol, shri\}@usc.edu,\\ \{simon.pistrosch, monica.gonzalez, schuller\}@tum.de} 

\usepackage{comment}
\usepackage{multirow,cite,bm}

\begin{document}

\maketitle

\begin{abstract}
    
We present a model for predicting articulatory features from surface electromyography (EMG) signals during speech production.
The proposed model integrates convolutional layers and a Transformer block, followed by separate predictors for articulatory features.
Our approach achieves a high prediction correlation of approximately 0.9 for most articulatory features.
Furthermore, we demonstrate that these predicted articulatory features can be decoded into intelligible speech waveforms. To our knowledge, this is the first method to decode speech waveforms from surface EMG via articulatory features, offering a novel approach to EMG-based speech synthesis. Additionally, we analyze the relationship between EMG electrode placement and articulatory feature predictability, providing knowledge-driven insights for optimizing EMG electrode configurations.
The source code and decoded speech samples are publicly available\footnote{\url{https://github.com/lee-jhwn/IS25-emg-ema}}.

\end{abstract}

\section{Introduction}

Methods for decoding speech from biosignals, such as electroencephalography, electrocorticography, magnetoencephalography, or electromyography, offer significant potential for advancing assistive technologies~\cite{gaddy-klein-2020-digital, imaginedspeech23, willett2023high, metzger2023high, proix2022imagined, fesde, fesde2, li2023neural2speech, meta-paper, gaddy2022voicing, hernaez2022,gonzalezlopez2020ssl}. These methods are particularly promising for facilitating speech production in individuals with communication impairments, by generating intelligible speech output from only partially available forms of speech, such as attempted or imagined articulation.

Compared to other biosignal recording techniques, surface electromyography (EMG) offers a practical and cost-effective alternative due to its non-invasive nature and lower device cost. It can be used to track muscle activities associated with speech production through electrodes placed on the skin surface of the face and neck.
Numerous studies have explored decoding speech from EMG signals, employing both manual extraction of EMG features~\cite{jou2006towards, schultz2010modeling, diener2016} and automatic extraction of features utilizing deep learning-based EMG representations~\cite{Li2022, gaddy2022voicing, scheck-2023, ren-2024}. Recent deep learning approaches primarily employ convolutional neural networks, recurrent neural networks, and Transformers~\cite{transformer} to predict speech units or intermediate acoustic features, such as mel-frequency cepstral coefficients or mel-spectrograms~\cite{diener2018, gaddy-klein-2020-digital, gaddy-klein-2021-improved, gaddy2022voicing, scheck-2023, ren-2024}. However, these approaches directly map EMG to the intermediate acoustic features or a latent space, making it difficult to interpret how muscle activities relate to speech and limiting controllability over speech output.

Recent advances in articulatory coding offer an alternative by enabling bidirectional transformation between speech waveforms and articulatory features obtainable through techniques such as electromagnetic articulography (EMA)~\cite{wu22i_interspeech, peter_aai_2023, cho2024coding}. For example, Cho et al.~\cite{cho2024coding} introduce a novel approach of two articulatory models, \textit{articulatory analysis} and \textit{articulatory synthesis} models, where EMA, pitch, and loudness are estimated from speech waveforms and vice versa with high accuracy. Another study proposes the integration of multiple articulatory modalities 
to enhance speech synthesis performance from each articulatory modality~\cite{wu2024deep}. A recent pilot study explores the feasibility of simultaneous recordings of surface EMG and EMA to enhance our understanding of the relationships between neuromotor and articulatory representations of speech~\cite{emg-ema-sync-2023}. 

These articulatory-based approaches improve interpretability and controllability by leveraging the physiological properties of speech production. However, a major limitation in applying articulatory synthesis to EMG-based speech decoding is the absence of sufficiently large synchronous EMG-EMA datasets to support deep learning approaches. Hence, we employ the recent development in articulatory coding to estimate articulatory trajectories from speech acoustics, thereby overcoming this limitation.

In this study, we propose a novel modeling framework that predicts articulatory features from surface EMG signals during vocalized speech production. Unlike previous EMG-based models that map to intermediate acoustic features or a latent space, our method decodes speech via articulatory features, providing greater interpretability.
We define \textit{articulatory features} as a combination of EMA (tongue, lip, and jaw movement), pitch (F0), and loudness (amplitude) features.
The proposed architecture employs an EMG encoder that processes input EMG signals, followed by separate predictors for EMA, pitch, loudness, and phoneme sequences.
The predicted articulatory features are then decoded into intelligible speech waveforms using an articulatory synthesis model \cite{cho2024coding}.
Additionally, we examine the relation between EMG electrode placement and predictability of articulatory movements, identifying key electrode locations that contribute most to accurate articulation modeling. The proposed model achieves an average correlation of 0.9 for EMA and loudness prediction.
Our results suggest that this knowledge-guided approach can help identify an optimal subset of EMG electrode configurations, reducing the number of required electrodes while maintaining predictive accuracy. 

A key advantage of our approach is articulatory interpretability, providing insights into the relationship between EMG signals and linguistically meaningful articulatory movements. By mapping the muscle activities recorded from the skin surface to articulatory actions, our method enhances the analysis of articulatory dynamics.

\begin{figure}[t]
  \centering
  \includegraphics[width=0.5\linewidth]{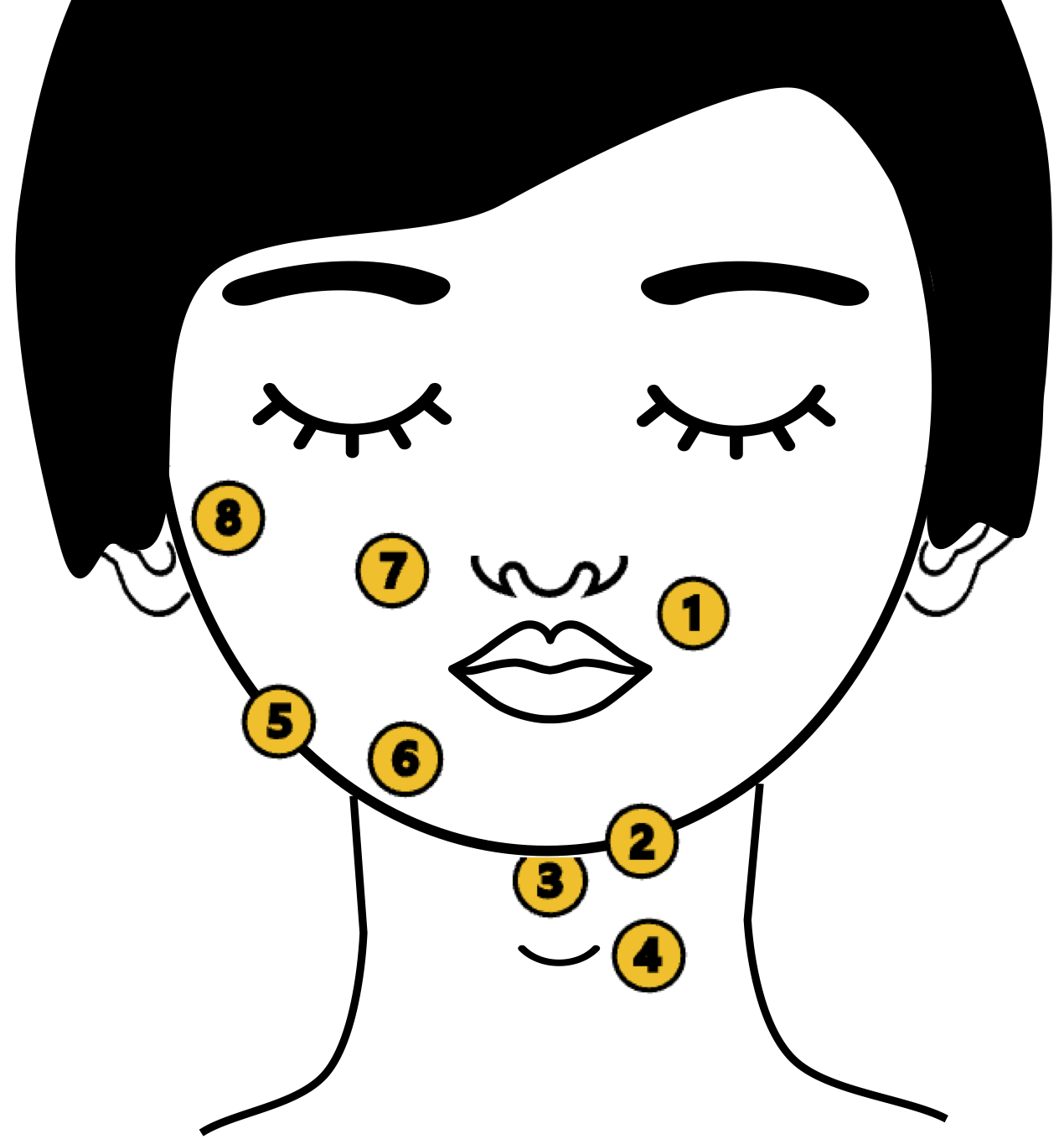}
  \caption{Visualization of EMG electrode placement.}
  \label{fig:emg-loc}
\end{figure}




\section{Methods}
\begin{table}[t]
\footnotesize
  \caption{Description of each EMG electrode location as in \cite{gaddy2022voicing}.}
  \label{tab:emg_loc}
  \centering
  \begin{tabular}{cc}
    \toprule
    \multicolumn{1}{c}{\textbf{EMG Electrode}} & \multicolumn{1}{c}{\textbf{Location}}\\
    \toprule

1 & left cheek just above mouth \\
2 & left corner of chin \\
3 & below chin back 3 cm \\
4 & throat 3 cm left from Adam’s apple \\
5 & mid-jaw right \\
6 & right cheek just below mouth \\
7 & right cheek 2 cm from nose \\
8 & back of right cheek, 4 cm in front of ear\\

    \bottomrule
  \end{tabular}
      \vspace{-4mm}
\end{table}

\subsection{Model Architecture}
\begin{figure}[t]
  \centering
  \includegraphics[width=0.55\linewidth]{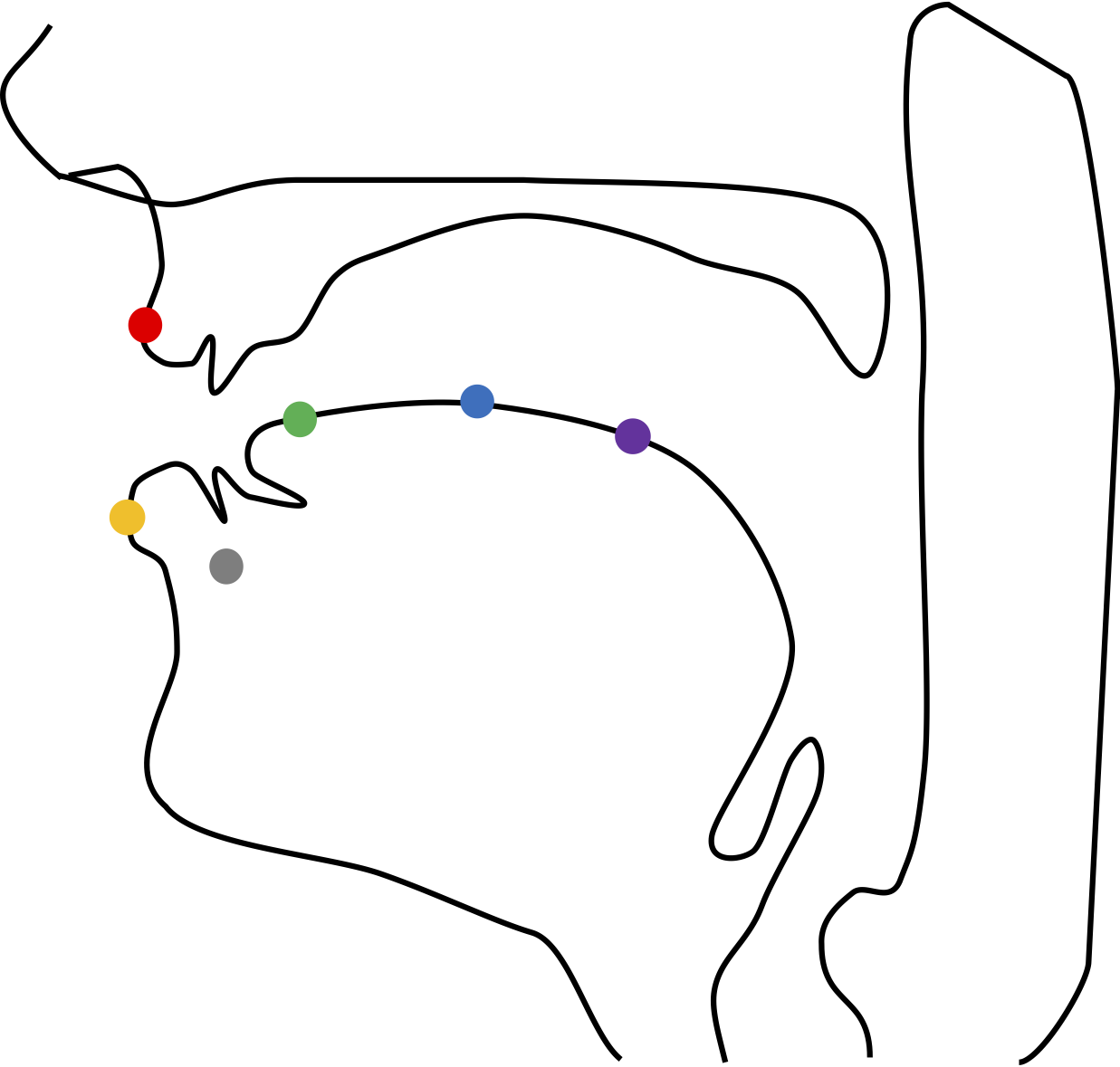}
  \caption{Six EMA sensor locations: Upper Lip (UL, red), Lower Lip (LL, yellow), Lower Incisor (LI, gray), Tongue Tip (TT, green), Tongue Body (TB, blue), and Tongue Dorsum (TD, purple).}
  \label{fig:ema-loc}
\end{figure}

\begin{figure}[t]
  \centering
  \includegraphics[width=\linewidth, trim={1.1cm 0.1cm 0.4cm 0},clip]{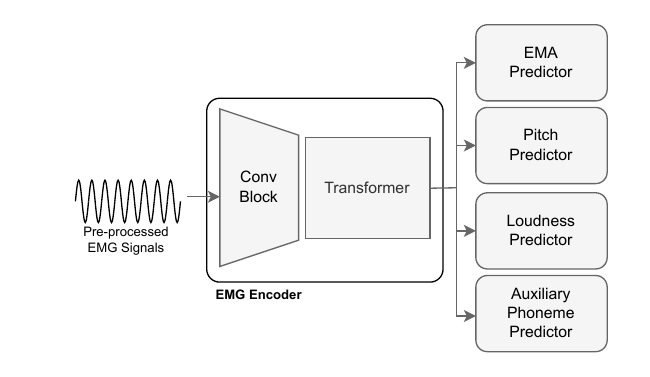}
  \caption{Overall architecture of the proposed framework. The EMG encoder processes input EMG signals using convolutional blocks and a six-layer Transformer block. Its output is fed into four predictors, EMA, pitch, loudness, and auxiliary phoneme predictors, each trained to estimate the corresponding articulatory feature.}
  \label{fig:overall-archi}
\end{figure}

As shown in Figure~\ref{fig:overall-archi}, the proposed model architecture consists of an EMG encoder that processes input EMG signals, followed by four predictors, each responsible for a specific output feature. 
The EMG encoder, adopting the approach in \cite{gaddy2022voicing}, comprises a convolutional block, followed by a six-layer Transformer \cite{transformer} block.
The convolutional block consists of three ResNet blocks, similar to \cite{he2016deep}, with convolutional layers having a kernel size of 3 and stride size of 2, along with ReLU activation and batch normalization.
The output of the EMG encoder is fed into four different predictors—EMA, pitch, loudness, and auxiliary phoneme predictors—each implemented as a single linear projection layer.
The convolutional block and Transformer have a hidden dimension of $768$.
We employ the auxiliary phoneme predictor as it stabilizes the training process in various biosignal-to-speech tasks \cite{gaddy2022voicing, fesde2}. For detailed model configuration, refer to \cite{gaddy2022voicing}.

\subsection{Training Objectives}
We adopt the L2 loss for EMA, pitch, and loudness prediction, and the cross-entropy loss for the auxiliary phoneme predictor per frame, following \cite{gaddy2022voicing}. The total loss is defined in Eq.(\ref{eq:total-loss}), and we heuristically set $\alpha_{\text{pitch}}$, $\alpha_{\text{loud}}$, and $\alpha_{\text{phon}}$ as $0.5$, $1.0$, and $0.5$, respectively.
\begin{align}
L_{total}=L_{\text{EMA}} + \alpha_{\text{pitch}} L_{\text{pitch}} + \alpha_{\text{loud}} L_{\text{loud}} + \alpha_{\text{phon}} L_{\text{phon}}
\label{eq:total-loss}
\end{align}

\section{Experiment Design}

\subsection{Dataset}
We conducted experiments using the dataset from \cite{gaddy-klein-2020-digital}, which consists of $7,565$ utterances, totaling $15.6$ hours of vocalized speech from a single male American English speaker.
The speech audio was recorded using a built-in laptop microphone at a sampling rate of $16,000\,Hz$.
EMG recordings were made from eight locations on the face and neck, shown in Table~\ref{tab:emg_loc} at a sampling rate of $1000\,Hz$.
The EMG recordings were pre-processed following \cite{gaddy2022voicing}, including powerline noise removal at $60\,Hz$, DC offset and drift removal using a high-pass filter at $2\,Hz$, soft de-spiking, then resampling from $1000\,Hz$ to $689\,Hz$. All other EMG set-up details remain identical to \cite{gaddy2022voicing}.


Despite recent efforts to collect synchronized EMA and EMG data~\cite{emg-ema-sync-2023}, no publicly available dataset is currently large enough to train a deep learning model.
To address this limitation, we estimated articulatory features from audio recordings using a recent acoustics-to-articulatory inversion (AAI) model~\cite{cho2024coding}. 
The estimated articulatory features include 12-dimensional EMA (three points from the tongue, two from the lips and one from the jaw, each with x- and y-axis values) as shown in Figure~\ref{fig:ema-loc}, pitch, and loudness.
The articulatory features were originally output at $50\,Hz$ and then resampled to $86.16\,Hz$ to match the output rate of the EMG encoder.
Additionally, the pitch values were normalized by subtracting $130$ 
then dividing by $70$, as this empirically improves the pitch prediction performance.

We adopted the identical train, validation, and test set splits as in \cite{gaddy2022voicing} with $200$ and $99$ utterances assigned to the validation and test sets, respectively.
Our experiments focused only on vocalized, open-vocabulary speech production.

\subsection{Experimental Setup}
All models were trained with the AdamW optimizer \cite{loshchilov2018decoupled} with a batch size of $32$, a learning rate of $0.0005$, and a weight decay of $10^{-7}$. The detailed training configuration is identical to \cite{gaddy2022voicing}. Model performance was measured at epoch 80, where validation loss indicated convergence.
All experiments were performed on a single Nvidia A40 GPU.

\begin{figure}[t]
  \centering
  \includegraphics[width=\linewidth]{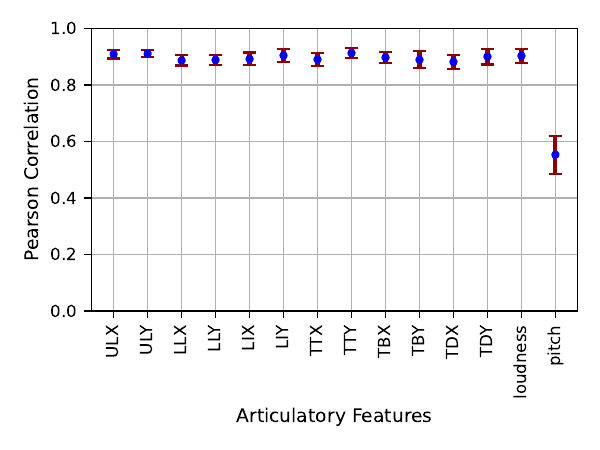}
  \caption{Pearson correlation between predicted articulatory features from EMG signals and the target articulatory features, estimated from audio recordings, along with 95\% confidence intervals.}
  \label{fig:arti-corr-all}
\end{figure}

\section{Results and Discussion}

\subsection{Articulatory Feature Prediction}

Figure~\ref{fig:arti-corr-all} shows the Pearson correlation between predicted and target articulatory features. The target features serve as pseudo-ground truths estimated from the corresponding acoustic signals following \cite{cho2024coding}. The model demonstrates strong predictive performance for EMA and loudness, yielding correlation values of approximately 0.9. For pitch prediction, the model demonstrates correlation values close to 0.6.

\subsection{Mapping EMG Electrode Activity to EMA Sensor Locations}

\begin{figure*}[t]
  \centering
  \subfloat[\textbf{Removing} one EMG electrode.]{
  \includegraphics[width=0.45\linewidth]{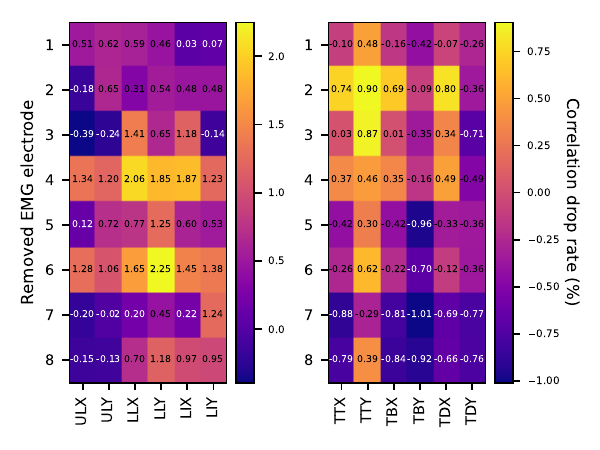}
  }
  \subfloat[\textbf{Using} only one EMG electrode.]{
    \includegraphics[width=0.45\linewidth]{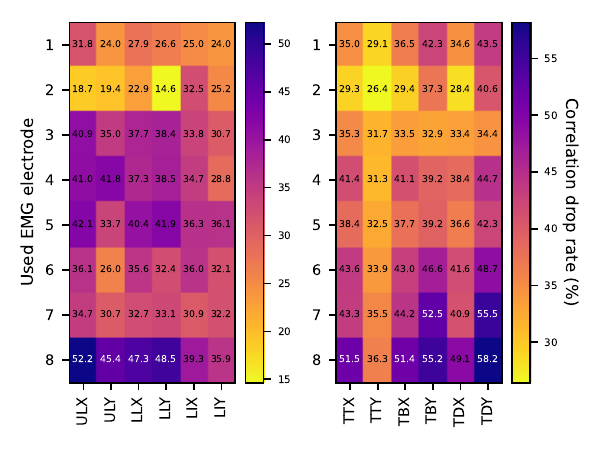}
  }
  \vspace{-1mm}
  \caption{Correlation drop rate for each EMA sensor location when one EMG electrode is \textbf{removed} (left) and when only one EMG electrode is \textbf{used} (right). Warmer (yellow) colors indicate a stronger association between the corresponding EMG electrode and EMA sensor location. Note that the color bars are inverted between the two figure panels to enhance readability.}
  \label{fig:emg-ema}
\end{figure*}

We investigate the contribution of each EMG electrode to EMA sensor prediction by measuring correlation drops under two conditions: (1) when a single EMG electrode is \textbf{removed}; and (2) when only a single EMG electrode is \textbf{used}.
The correlation drop rate is computed relative to the case where all eight EMG electrodes are used, as shown in Figure~\ref{fig:emg-ema}. In both plots, warmer colors indicate a stronger association between each EMG electrode and the corresponding EMA sensor positions.

In general, we observe stronger association in the lips compared to the tongue. This may be attributed to the nature of EMG sensors, which rely on surface measurements. Since the lips are externally positioned, their movements are directly captured by the EMG sensors. In contrast, the tongue, being internally located, poses greater challenges for accurate movement measurement using external skin surface EMG sensors.

Regarding tongue movement, EMG electrode 2 exhibits the strongest effect, particularly on the tongue tip position. Located at the corner of the chin, just superficial to the attachments of the genioglossus fibers, electrode 2 likely picks up activities from the anterior genioglossus, which is responsible for advancing and raising of the front part of the tongue. EMG electrode 3, placed in the posterior submental region and proximal to the genioglossus, also shows a stronger association with tongue movement.
For lower lip and jaw movement, EMG electrodes 2 and 6, both show a strong association, indicating their relevance to labial sounds. This relationship may be attributed to their proximity to muscles involved in raising and lowering the jaw and lower lip, allowing them to effectively capture muscle activity associated with articulatory movements in this region.


EMG electrode 4, located on the neck near the paralaryngeal region, exhibits strong association with most articulatory features, including EMA, pitch, and loudness. Using this electrode alone results in a correlation of $0.73$ for loudness prediction and $0.32$ for pitch prediction, the highest among all EMG electrodes. Note that this electrode is positioned closest to the larynx, the primary site of voicing.



\subsection{Intelligibility of Decoded Speech}
\begin{table}[t]
\scriptsize
  \caption{Phoneme error rate (PER), word error rate (WER), and SpeechBERTScore (SBS), with 95\% confidence intervals.}
  \label{tab:per-wer}
  \vspace{-0.2cm}
  \centering
  \begin{tabular}{cccc}
    \toprule
    \multicolumn{1}{c}{\multirow{1}{*}{\textbf{Model}}} & \multicolumn{1}{c}{\textbf{PER (\%) $\downarrow$}}& \multicolumn{1}{c}{\textbf{WER (\%) $\downarrow$}} & \multicolumn{1}{c}{\textbf{SBS (\%)$\uparrow$}}\\
    \toprule
    \textsc{Ground truth}
    & $30.1 \pm 5.3$ & $8.6 \pm 3.7$ & $100$\\
    \textsc{Resynthesis} &$32.9 \pm 4.7$&$10.6 \pm 3.7$&$86.1 \pm 0.9$\\
    \textsc{Gaddy et al. \cite{gaddy2022voicing}} &$33.9 \pm 3.1$ & $12.6 \pm 2.2$ & $77.9 \pm 0.8$\\
    \textsc{Ours} & $36.0 \pm 2.7$ & $15.5 \pm 3.5$ & $84.4 \pm 0.7$\\


    \bottomrule
  \end{tabular}
  
\end{table}
\begin{table*}[ht]
\footnotesize
  \caption{Performance when using a subset of EMG electrodes. Pearson correlation of predicted articulatory features (EMA, loudness, and pitch), phoneme error rate (PER), word error rate (WER), and SpeechBERTScore (SBS), with 95\% confidence intervals.}
  \label{tab:subchannel}
  \vspace{-0.2cm}
  \centering
  \begin{tabular}{cc|ccc|ccc}
    \toprule
    \multicolumn{2}{c|}{\multirow{2}{*}{\textbf{EMG electrodes}}} & \multicolumn{3}{c|}{\textbf{Correlation of predicted articulatory features}} & \multicolumn{3}{c}{\textbf{Evaluation on integellibility}} \\
    &&\textbf{EMA} & \textbf{Loudness} & \textbf{Pitch} &\multicolumn{1}{c}{\textbf{PER (\%) $\downarrow$}}& \multicolumn{1}{c}{\textbf{WER (\%) $\downarrow$}} & \multicolumn{1}{c}{\textbf{SBS (\%)$\uparrow$}}\\
    \toprule
    \textsc{Set of 4} & ch.2,3,4,6
    & $0.882 \pm 0.017$ & $0.904 \pm 0.057$ & $0.538 \pm 0.342$ & $39.7 \pm 5.2$ & $22.9 \pm 3.7$ & $83.2 \pm 0.7$\\
        \textsc{Set of 3} & ch.2,4,6
    & $0.859 \pm 0.021$ & $0.892 \pm 0.048$ & $0.562 \pm 0.343$ & $46.1 \pm 6.0$ & $31.6 \pm 3.9$ & $82.1 \pm 0.8$\\
    \textsc{Set of 2} & ch.2,4
    & $0.816 \pm 0.023$ & $0.870 \pm 0.078$ & $0.576 \pm 0.286$ & $52.1 \pm 4.6$ & $46.9 \pm 5.1$ & $79.9 \pm 0.7$\\


    \bottomrule
  \end{tabular}
  \vspace{-5mm}
\end{table*}

We examine whether predicted articulatory features from EMG input can be further decoded into intelligible speech waveforms.
To assess intelligibility objectively, we measure phoneme error rate (PER), word error rate (WER), and SpeechBERTScore (SBS) \cite{saeki2024speechbertscore}. We compute PER by comparing the phoneme transcriptions from Montreal forced aligner (MFA)~\cite{mfa}, accompanying the dataset, with the phoneme sequence predicted by a Wav2Vec2-XLSR model fine-tuned for phoneme recognition~\cite{xu2021simpleeffectivezeroshotcrosslingual}. WER is measured by comparing word transcription from the same MFA alignments with text predictions from whisper-large-v3~\cite{radford2022robustspeechrecognitionlargescale}.
SpeechBERTScore~\cite{saeki2024speechbertscore} is a recently introduced metric that quantifies the similarity between the reference and generated speech samples by leveraging self-supervised learning. It has been shown to highly correlate with human evaluation of speech synthesis quality. We set the ground-truth speech waveforms from the recordings as the reference.
\textsc{Ground truth} indicates the actual speech recordings and \textsc{Resynthesis} refers to direct reconstruction of speech waveforms from the target articulatory features that are estimated from ground-truth speech waveforms, serving as the upper bound in our model. In \textsc{Gaddy et al.~\cite{gaddy2022voicing}}, the speech waveforms are decoded via mel-spectrogram prediction by the HiFi-GAN vocoder \cite{kong2020hifi}, fine-tuned on this dataset.
In the proposed model~\textsc{Ours}, the predicted articulatory features are converted into speech waveforms using the articulatory synthesis model from \cite{cho2024coding}.

As shown in Table~\ref{tab:per-wer}, our proposed approach via articulatory features exhibits intelligibility comparable to \textsc{Gaddy et al.~\cite{gaddy2022voicing}} in terms of PER and WER, though slightly worse. Notably, for SpeechBERTScore, our method outperforms the previous approach based on mel-spectrogram \cite{gaddy2022voicing}, and approaches the upper bound of \textsc{Resynthesis} very closely.
These results demonstrate that our proposed method achieves comparable performance to the previous approaches while offering greater articulatory interpretability. 



\subsection{Ablation Study: Articulatory Feature-based Selection of EMG Electrodes}

Reducing the number of EMG electrodes improves both practicality and cost-effectiveness. However, previous studies \cite{gaddy2022voicing} have relied on random search to achieve this, which is computationally expensive and requires extensive experiments.
Based on the contribution of each EMG electrode to different articulatory features, we pilot to utilize this knowledge to select a smaller subset of EMG electrodes with minimal performance degradation.
We select the four most influential EMG electrodes: 2, 3, 4, and 6. Electrodes 2 and 3 contribute most to tongue movement, while electrodes 4 and 6 are strongly associated with lower lip and jaw movement. Additionally, electrode 4, located in the paralaryngeal region, near the primary site of vocal fold vibration, shows the most contribution to pitch and loudness prediction.

We compare the performance of this 4-electrode subset along with even smaller electrode configurations.
As shown in Table~\ref{tab:subchannel}, using these four EMG electrodes still allows the model to function effectively\footnote{The decoded speech samples from different subsets of EMG electrodes are available on the sample page.}
This aligns with \cite{gaddy2022voicing}'s finding, where this specific set of EMG electrodes is also identified as a well-performing subset. More importantly, our articulatory feature-based approach offers a principled framework for optimizing electrode placement based on articulatory relevance, providing a more efficient and informed alternative to random search.


\subsection{Limitations}
Our approach has certain limitations. The articulatory features, particularly EMA, are estimated from speech recordings rather than directly measured from EMA sensors from actual recordings. This is due to the lack of a publicly available parallel EMG-EMA dataset of sufficient size to train a deep learning model.
Consequently, our approach is inherently influenced by the accuracy of the EMA estimation model \cite{cho2024coding}.
Additionally, the current model is constrained to a single speaker and does not generate the speaker embedding required for the articulatory synthesis model \cite{cho2024coding}. Instead, the speaker embedding is extracted from an external model. Addressing these limitations in future work could further enhance the generalizability and robustness of EMG-based articulatory feature prediction and speech synthesis.

\section{Conclusion}

In this paper, we propose to predict articulatory features from surface EMG signals during speech production. We also show that the predicted articulatory speech waveforms can be further decoded to intelligible speech waveforms. A deeper analysis on the relationships between each EMG electrode and the articulatory features is also provided. We further demonstrate a knowledge-driven exploration for selecting a smaller set of EMG electrodes.


\section{Acknowledgment}

This work was supported by the National Science Foundation. (IIS-2311676, BCS-2240349)

\bibliographystyle{IEEEtran}

\bibliography{mybib}

\end{document}